\documentclass[letter,twocolumn]{jpsj3}
\usepackage{amsmath}
\usepackage{amssymb}
\usepackage{graphicx}
\usepackage{color}

\title{
Interplay of Spin-Orbit Interaction and Electron Correlation on the Van Vleck Susceptibility in Transition Metal Compounds
}

\author{Masafumi \textsc{Udagawa} and Youichi \textsc{Yanase}$^{1}$}

\inst{
Department of Applied Physics, School of engineering, University of Tokyo, Hongo, Tokyo 113-8656, Japan\\
$^{1}$ Department of Physics, Niigata University, Niigata 950-2181, Japan
}

\recdate{\today}

\abst{
We have studied the effects of electron correlation on Van Vleck susceptibility ($\chi_{\rm{VV}}$) in transition metal compounds. A typical crossover behavior is found for the correlation effect on $\chi_{\rm{VV}}$ as sweeping spin-orbit interaction, $\lambda$. For a small $\lambda$, orbital fluctuation plays a dominant role in the correlation enhancement of $\chi_{\rm{VV}}$; however, the enhancement rate is rather small. 
In contrast, for an intermediate $\lambda$, $\chi_{\rm{VV}}$ shows a substantial increase, 
accompanied by the development of spin fluctuation. We will discuss the behavior of $\chi_{\rm{VV}}$ in association 
with the results of Knight-shift experiments on Sr$_2$RuO$_4$ and an anomalously large magnetic susceptibility observed for $5d$ Ir compounds.
}

\kword{
Van Vleck susceptibility, multi-orbital system, spin-orbit interaction, dynamical mean-field theory
}

\begin{document}
\maketitle
Multi-orbital systems often show intriguing phenomena, 
such as unconventional superconductivity\cite{Maeno,Takada} and colossal magneto-resistance\cite{Tokura}.
In most cases, orbitals do not just complicate a system by increasing the number of local degrees of freedom, 
but they lead to qualitatively new features, which are absent in single-orbital systems. Among the characteristic properties inherent in multi-orbital systems, 
the role of orbital angular momentum and the spin-orbit coupling deserves special attention, 
since they underlie many anomalous phenomena in transition metal and rare-earth materials.

Above all, these two factors give rise to ``residual" paramagnetic susceptibility, called Van Vleck susceptibility ($\chi_{\rm{VV}}$)\cite{VanVleck,KuboObata}. It is well known that the Pauli susceptibility ($\chi_{\rm{P}}$) is proportional to the density of states,
and it vanishes, as soon as an energy gap opens at the Fermi level.
Meanwhile, $\chi_{\rm{VV}}$ has a residual nature, in the sense that it remains finite at zero temperature, even in the presence of an energy gap.

To be more specific, the residual nature of $\chi_{\rm{VV}}$ 
can be attributed to the fact that the orbital angular momentum or spin-orbit interaction makes magnetization
a non-conserved quantity.
To illustrate this, suppose a system in which the magnetization is conserved.
In this case, the Hamiltonian and magnetization operators are commutative, and thus a fixed magnetization quantum number can be assigned to each eigenstate of the Hamiltonian.
Consequently, for the magnetic susceptibility to be finite at zero temperature, the ground state has
to be replaced with one of the excited states with an infinitesimal magnetic field, which is obviously
impossible in a gapped system.
Meanwhile, for a system where the magnetization is not conserved, an infinitesimal magnetic field leads to magnetization by mixing the ground-state and excited-state wave functions,
resulting in a finite susceptibility even in the presence of an energy gap.

In fact, the residual nature of $\chi_{\rm{VV}}$ causes much confusion in gapped systems.
A typical example can be found in a Knight-shift experiment to determine the parity of a superconducting order parameter. Below the superconducting transition temperature $T_c$, only $\chi_{\rm P}$ decreases, responding to spin gap formation, while $\chi_{\rm{VV}}$ remains invariant. 
This means that the invariant Knight shift at $T_c$ does not necessarily serve as evidence of spin-triplet superconductivity, 
but suggests another possibility that a very large $\chi_{\rm{VV}}$ masks the decrease in $\chi_{\rm{P}}$ to experimental accuracy.
The most famous material that suffers from this difficulty may be UPt$_3$, 
for which a decrease in Knight shift has been observed; 
however, the decrease is only 1\% of the total Knight shift\cite{Tou}.
Similar difficulties have been reported for conventional vanadium-based superconductors\cite{Clogston}. Recent Knight-shift experiments on Sr$_2$RuO$_4$ also led to the same question. 
Murakawa and coworkers found that the magnetic susceptibility shows no change at $T_c$, irrespective of 
the magnetic field direction\cite{Murakawa,Murakawa2}.
In order to provide a basis for discussing these interesting superconductors, it is desirable to clarify the Van Vleck
susceptibility of multi-orbital systems.

$5d$ Ir compounds provide another example in which Van Vleck susceptibility plays a crucial role.
One of the confusing properties common in Ir compounds is their unusually large magnetic susceptibility and Wilson ratio.
For example, for Eu$_2$Ir$_2$O$_7$, the magnetic susceptibility amounts to $\chi\sim 1.0\times 10^{-2}$emu$/$mol-Ir, 
in contrast to the rather small specific heat coefficient $\gamma\sim 8.0$mJ/K$^2$ mol-Ir, leading to an anomalously large
Wilson ratio, $R_{\rm W}\sim 90$\cite{Yanagishima01}. A similar huge paramagnetic susceptibility has also been reported for the so-called ``hyperkagome material" Na$_4$Ir$_3$O$_8$, where a large residual magnetic susceptibility, $\chi\sim 1.0\times 10^{-3}$emu$/$mol-Ir was observed\cite{Okamoto}, despite that this compound is an insulator.
To reconcile a large paramagnetic susceptibility with a small $\gamma$, substantial increase in $\chi_{\rm{VV}}$ is necessary,
since a large $\chi_{\rm{P}}$ requires the existence of rich gapless spin excitations, which should also contribute to $\gamma$. 

Actually, in rare-earth systems, correlation effects on Van Vleck susceptibility have been studied by 
several groups\cite{ZouAnderson,ZhangLee,Kontani1,Mutou,Saso,Kontani2}. 
Among them, Kontani and Yamada pointed out the general tendency that the correlation enhancement of $\chi_{\rm{VV}}$ is 
comparable to that of $\chi_{\rm{P}}$\cite{Kontani1}.
However, it has also been reported that the enhancement rate considerably depends on individual properties of a system, such as orbital degeneracy\cite{Kontani2,Mutou}.
Therefore, it is highly non-trivial how electron correlation affects $\chi_{\rm{VV}}$ in transition metal compounds,
which have quite different characters from rare-earth materials.

In this research, we will study the correlation effect on $\chi_{\rm{VV}}$ in transition metal compounds. For this purpose, we adopt Sr$_2$RuO$_4$ as a model material, 
since its simple orbital structure is appropriate for establishing a general theory,
and the $\chi_{\rm{VV}}$ of Sr$_2$RuO$_4$ is interesting of its own right.
Although the spin-orbit coupling of Sr$_2$RuO$_4$ is rather small, 
we also investigate the case with a large spin-orbit coupling, in order to gain insights into large $\chi_{\rm{VV}}$ in {\it 5d} transition metal compounds. Hereafter, we set $\hbar=k_B=\mu_B=1$.

We start with a multi-orbital Hubbard model, which takes account of the
three $t_{2g}$ orbitals of Sr$_2$RuO$_4$,
\begin{eqnarray}
H = H_0 + H_I,
\label{Ham}
\end{eqnarray}
\begin{align}
&H_0 = \sum\limits_{{\mathbf k},s=\pm 1}
\begin{pmatrix}
c^{\dag}_{{\mathbf k},1,s} & c^{\dag}_{{\mathbf k},2,s} & c^{\dag}_{{\mathbf k},3,-s}
\end{pmatrix}\nonumber\\
 &\times 
 \begin{pmatrix}
\epsilon_1({\mathbf k}) & -\lambda s & -\lambda s \\
-\lambda s & \epsilon_2({\mathbf k}) & \lambda \\
-\lambda s & \lambda & \epsilon_3({\mathbf k})
 \end{pmatrix}
 \begin{pmatrix}
c_{{\mathbf k},1,s}\\
c_{{\mathbf k},2,s}\\
c_{{\mathbf k},3,-s}
\end{pmatrix},
\label{H0}
\end{align}
\begin{align}
&H_I = U\sum\limits_i\sum\limits_an_{i,a,\uparrow}n_{i,a,\downarrow} + U'\sum\limits_i\sum\limits_{a>b}n_{i,a}n_{i,b} \nonumber\\
& - J\sum\limits_i\sum\limits_{a\not=b}\Bigl[\Bigl({\mathbf S}_{i,a}{\mathbf S}_{i,b} + \frac{1}{4}n_{i,a}n_{i,b}\Bigl) -\zeta_{ab}c^{\dag}_{i,a,\downarrow}c^{\dag}_{i,a,\uparrow}c_{i,b,\uparrow}c_{i,b,\downarrow}\Bigl]\nonumber\\
&\equiv \sum\limits_i\sum\limits_{\eta_1\eta_2\eta_3\eta_4}I_{\eta_1\eta_2\eta_3\eta_4}c^{\dag}_{i\eta_1}c^{\dag}_{i\eta_2}c_{i\eta_3}c_{i\eta_4}.
\label{HI}
\end{align}
Here, $s=+1(-1)$ denotes the up- (down-) spin.
The indices $a,b=1,2,$ and $3$ represent $d_{yz}$, $d_{zx}$, and $d_{xy}$ orbitals, respectively.
Here, we introduce the abbreviation $\eta = (s, a)$.
We define $|yz\rangle=\frac{i}{\sqrt{2}}(|+1\rangle+|-1\rangle)$, 
$|zx\rangle=-\frac{i}{\sqrt{2}}(|+1\rangle-|-1\rangle)$, 
$|xy\rangle=-\frac{i}{\sqrt{2}}(|+2\rangle-|-2\rangle)$, where $|m\rangle$ is the eigenstate of the 
orbital angular momentum, with $l^z|m\rangle = m|m\rangle$. 
This convention makes $H_0$ real, at the cost of the sign factor in pair hopping terms: $\zeta_{ab}=-1$, if $a=2$ or $b=2$, and $\zeta_{ab}=1$, otherwise.
We determine the kinetic energy terms in $H_0$, by the two-dimensional tight-binding model: 
$\epsilon_1({\mathbf k}) = -2t'_{xy}\cos k_x - 2t_{xy}\cos k_y - \mu$, $\epsilon_2({\mathbf k}) = -2t_{xy}\cos k_x 
- 2t'_{xy}\cos k_y - \mu$ and $\epsilon_3({\mathbf k}) = -2t_z(\cos k_x + \cos k_y) - 4t'_z\cos k_x\cos k_y - \mu$.
We set $t_z=1$ as a unit of energy, and $t_{xy} = 1.5$, $t'_{xy} = 0.2$, $t'_z=0.4$ to reproduce the Fermi surface obtained
in the de Haas-van Alphen experiment\cite{Mackenzie}. 
The chemical potential $\mu$ is controlled so that the system is at $2/3$ filling.

In order to separate $\chi_{\rm{P}}$ and $\chi_{\rm{VV}}$, we follow the discussion by Kontani and Yamada\cite{Kontani1}.
We fix the magnetic field parallel to $z$; then, we define 
\begin{eqnarray}
\chi_{\rm{VV}} \equiv \lim_{\omega\to 0}\lim_{{\mathbf q}\to 0}\chi_{{\mathbf q}}^{zz}(\omega),
\label{chi_VV}
\end{eqnarray}
\begin{eqnarray}
\chi\equiv\chi_{\rm{P}} + \chi_{\rm{VV}}\equiv \lim_{{\mathbf q}\to 0}\lim_{\omega\to 0}\chi_{{\mathbf q}}^{zz}(\omega),
\end{eqnarray}
\begin{eqnarray}
\chi_{\rm{P}}\equiv\chi - \chi_{\rm{VV}},
\label{chi_P}
\end{eqnarray}
where the magnetic correlation function $\chi_{{\mathbf q}}^{\alpha\alpha'}(\omega)$ can be obtained from its Matsubara-frequency representation
$\chi^{\alpha\alpha'}_{{\mathbf q}}(i\omega_q) = \frac{1}{N}\int\limits_0^{\beta}d\tau\ e^{i\omega_q\tau}
\langle M^{\alpha}_{-{\mathbf q}}(\tau)M^{\alpha'}_{{\mathbf q}}\rangle$ ($\alpha,\alpha'=x,y,z$) as $\chi^{\alpha\alpha'}_{{\mathbf q}}(\omega) = \lim_{\beta\to\infty}\chi^{\alpha\alpha'}_{{\mathbf q}}(i\omega_q\rightarrow\omega + i\delta)$, where $M^{\alpha}_{{\mathbf q}} =\sum_{{\mathbf k},\eta\eta'}\langle\eta|
M^{\alpha}|\eta'\rangle c^{\dag}_{{\mathbf k}+{\mathbf q},\eta}c_{{\mathbf k},\eta'}$,
with $M^{\alpha} = l^{\alpha}+2s^{\alpha}$, and $N$ is the number of sites.
The definitions eqs. (\ref{chi_VV})-(\ref{chi_P}) guarantee the properties required for
Pauli and Van Vleck susceptibilities: $\chi_{\rm{P}}$ vanishes if an infinitesimal energy gap opens at the Fermi level.
These definitions are meaningful only at zero temperature. However, for numerical calculation, we introduce a small temperature, $T$, and approximate $\chi_{\rm{VV}}$ as
$\chi_{\rm{VV}} \simeq \chi_{{\mathbf q}=0}^{zz}(i\pi T) - \frac{\chi_{{\mathbf q}=0}^{zz}(3i\pi T) - \chi_{{\mathbf q}=0}^{zz}(i\pi T)}{2}$.
We typically set $T=0.01$.

In order to obtain $\chi_{{\mathbf q}}(i\omega_q)$, we use the dynamical mean-field theory (DMFT)\cite{Georges}, in which
the self-energy and the irreducible vertex function are approximated by the local ones.
In our model, the local Green's function can be expressed as a $6\times 6$ matrix, $G_{\eta\eta'}(i\epsilon_p)$, which satisfies the self-consistent equations involving the
Weiss field, $g_{\eta\eta'}(i\epsilon_p)$.\cite{Georges} Here, we obtain the self-energy $\Sigma_{\eta\eta'}(i\epsilon_p)$ using the iterative perturbation theory (IPT): we expand the self-energy to the
second order of electron interactions, $U$, $U'$, and $J$. This method is valid only for a weak-coupling region, however, 
computationally inexpensive and appropriate for studying a wide parameter range.
The self-energy can be written as
\begin{align}
&\Sigma_{\eta_1\eta_2}(\tau) = \Gamma^{(0)}_{\eta_1\eta_2\eta_3\eta_4}g_{\eta_3\eta_2}(\tau=0-)\nonumber\\
 &+ \frac{1}{2}\Gamma^{(0)}_{\eta_1\eta'_2\eta'_3\eta'_4}\Gamma^{(0)}_{\eta''_1\eta''_2\eta''_3\eta_2}g_{\eta'_3\eta''_1}(\tau)g_{\eta'_4\eta''_2}(\tau)g_{\eta''_3\eta'_2}(-\tau),
\end{align}
where we have introduced the bare vertex function
$\Gamma^{(0)}_{\eta_1\eta_2\eta_3\eta_4} = 2(I_{\eta_1\eta_2\eta_3\eta_4} - I_{\eta_2\eta_1\eta_3\eta_4})$.\cite{omitFock}
Then the magnetic correlation function is written as
\begin{align}
\chi^{\alpha\alpha'}_{{\mathbf q}}(i\omega_q) &= \sum\limits_{\eta_1\eta_2\eta_3\eta_4}T\sum\limits_{\epsilon_p}T\sum\limits_{\epsilon_p'}G_{{\mathbf q},\eta_1\eta_2\eta_3\eta_4}(i\omega_q;i\epsilon_p,i\epsilon_p')\nonumber\\
&\times \langle\eta_3|M^{\alpha}|\eta_1\rangle\langle\eta_4|M^{\alpha'}|\eta_2\rangle e^{i\epsilon_p0+}e^{i\epsilon_p'0+},
\end{align}
where the two-body Green's function $G_{{\mathbf q},\eta_1\eta_2\eta_3\eta_4}$ can be obtained from the Bethe-Salpeter equation\cite{Georges}, in which 
we calculate the local irreducible vertex function up to the second order of electron interactions as
\begin{align}
&{\Gamma}_{\eta_1\eta_2\eta_3\eta_4}(i\omega_q;i\epsilon_p,i\epsilon_p')\nonumber\\
&= \Gamma^{(0)}_{\eta_1\eta_2\eta_3\eta_4} - \Gamma^{(0)}_{\eta_1\eta'_2\eta'_3\eta_4}\Gamma^{(0)}_{\eta'_1\eta_2\eta_3\eta'_4}T^{+}_{\eta'_4\eta'_2\eta'_3\eta'_1}(i(\epsilon_p-\epsilon_p'))\nonumber\\
&+ \frac{1}{2}\Gamma^{(0)}_{\eta_1\eta_2\eta'_3\eta'_4}\Gamma^{(0)}_{\eta'_1\eta'_2\eta_3\eta_4}T^{-}_{\eta'_3\eta'_2\eta'_4\eta'_1}(i(\epsilon_p+\epsilon_{p'}+\omega_q)), 
\end{align}
and $T^{\pm}_{\eta_1\eta_2\eta_3\eta_4}(i\omega_q)=-T\sum\limits_{\epsilon_p}G_{\eta_1\eta_2}(i\epsilon_p)G_{\eta_3\eta_4}(\pm i\epsilon_p+i\omega_q)$.

\begin{figure}[h]
\begin{center}
\includegraphics[width=0.45\textwidth]{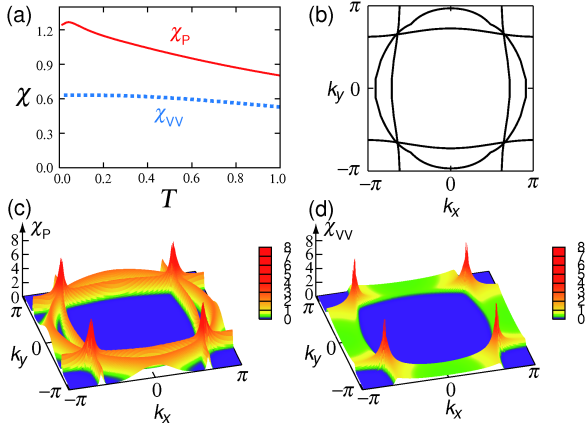}
\end{center}
\caption{\label{U0} 
(Color online) 
(a) Temperature dependences of $\chi_{\rm{P}}$ and $\chi_{\rm{VV}}$ in non-interacting case and $\lambda=0$. 
(b) Fermi surface for $\lambda=0$. (c) $\chi_{\rm{P}}({\mathbf k})$ and (d) $\chi_{\rm{VV}}({\mathbf k})$.
}
\end{figure}

Below, we show our results.
Firstly, we will discuss $\chi_{\rm{P}}$ and $\chi_{\rm{VV}}$ for a non-interacting case and $\lambda=0$.
In this case, the definitions of $\chi_{\rm{P}}$ and $\chi_{\rm{VV}}$, eqs. (\ref{chi_VV})-(\ref{chi_P}) can
be naturally extended to a finite temperature. We add the Zeeman term $\mathcal{H}_Z=-h\sum_{i,\eta\eta'}\langle\eta|M^z|\eta'\rangle c^{\dag}_{i\eta}c_{i\eta'}$ to the Hamiltonian eq. (\ref{Ham}) and obtain
$\chi_{\rm{P}}= \frac{1}{N}\sum_{\mathbf k} \chi_{\rm{P}}({\mathbf k})= \frac{1}{N}\sum_{{\mathbf k},a}\bigl(\frac{\partial\epsilon_{a}({\mathbf k})}{\partial h}\bigr)^2\bigl(-\frac{\partial}{\partial\epsilon}f(\epsilon_{a}({\mathbf k}) - \mu)\bigr)$, and
$\chi_{\rm{VV}}= \frac{1}{N}\sum_{\mathbf k} \chi_{\rm{VV}}({\mathbf k})= -\frac{1}{N}\sum_{{\mathbf k}, a}\bigl(\frac{\partial^2\epsilon_{a}({\mathbf k})}{\partial h^2}\bigr)f(\epsilon_{a}({\mathbf k}) - \mu)$,
with a Fermi distribution function, $f(x)=\frac{1}{e^{\beta x}+1}$.
We plot the temperature dependences of $\chi_{\rm{P}}$ and $\chi_{\rm{VV}}$ in Fig. \ref{U0} (a).
$\chi_{\rm{P}}$ shows a moderate increase with decreasing temperature, in contrast to 
$\chi_{\rm{VV}}$, which takes almost a constant value in a wide temperature range, $0\lesssim T\lesssim 1$.

Figures \ref{U0} (c) and \ref{U0} (d) show the momentum-resolved magnetic susceptibilities $\chi_{\rm{P}}({\mathbf k})$ and $\chi_{\rm{VV}}({\mathbf k})$ evaluated at $T=0.1$, respectively. The major contribution to $\chi_{\rm{P}}$ comes from 
the vicinity of three Fermi surface sheets, while $\chi_{\rm{VV}}$ comes from a wide area in a Brillouin zone (e.g., around ($\pi, 0$) and ($0, \pi$)
), where either the $d_{yz}$ or $d_{zx}$ orbital is occupied, and the other orbital is empty.
We note that $\chi_{\rm{VV}}$ is brought about by the hybridization of the $d_{yz}$ and $d_{zx}$ orbitals due to the magnetic field parallel to the $z$-axis.
In particular, sharp peaks are located at $(p_x, p_y)\sim(\pm0.62\pi, \pm0.62\pi)$, where these bands cross at the Fermi level.

\begin{figure}[h]
\begin{center}
\includegraphics[width=0.45\textwidth]{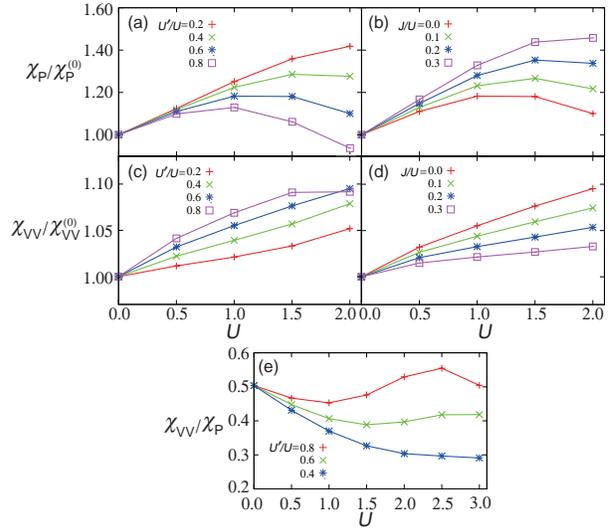}
\end{center}
\caption{\label{Lambda0} 
(Color online) 
(a) (b) $U$ dependence of $\chi_{\rm{P}}/\chi_{\rm{P}}^{(0)}$. (c) (d) The $U$ dependence of $\chi_{\rm{VV}}/\chi_{\rm{VV}}^{(0)}$. 
We fix $J/U=0.0$ for (a) and (c), and $U'/U=0.6$ for (b) and (d). (e) Ratio of $\chi_{\rm{P}}$ to $\chi_{\rm{VV}}$ for the
interaction parameters under the relation $U'=U-2J$. We fix $\lambda=0$ in (a)-(e). 
}
\end{figure}

Next, let us consider the effect of electron interaction.
In Fig. \ref{Lambda0}, we show the $\chi_{\rm{P}}$ and $\chi_{\rm{VV}}$ divided by their non-interacting values $\chi_{\rm{P}}^{(0)}$ and $\chi_{\rm{VV}}^{(0)}$.
Here, we plot $\chi_{\rm{P}}$ and $\chi_{\rm{VV}}$ by varying $U$, with $U'/U$ and $J/U$ fixed.
Figures \ref{Lambda0}(a)-\ref{Lambda0}(d) show that both $\chi_{\rm{P}}$ and $\chi_{\rm{VV}}$ tend to increase with $U$; however, their growth rates are quite different. Although $\chi_{\rm{P}}$ is substantially enhanced by electron interaction,
$\chi_{\rm{VV}}/\chi_{\rm{VV}}^{(0)}$ remains $\sim 1.1$, at most, in the interaction range considered here.
Moreover, Figs. \ref{Lambda0} clearly show that the inter-orbital repulsion $U'$ and Hund coupling $J$ affect
$\chi_{\rm{P}}$ and $\chi_{\rm{VV}}$, quite differently.
Figures \ref{Lambda0}(a) and \ref{Lambda0}(b) show that $\chi_{\rm{P}}$ monotonically increases with $J$, 
while it is suppressed with increasing $U'$.
In contrast, Figs. \ref{Lambda0}(c) and \ref{Lambda0}(d) show that $\chi_{\rm{VV}}$ decreases with $J$, 
while it grows with increasing $U'$. 
The contrastive behaviors of $\chi_{\rm{P}}$ and $\chi_{\rm{VV}}$ can be attributed to the difference in the way
electron correlation affects spin and orbital fluctuations.
In multi-orbital systems, a magnetic moment can be decomposed into the spin part and the orbital part, as $M^z=l^z + 2s^z$.
Without spin-orbit coupling, $\chi_{\rm{P}}$ ($\chi_{\rm{VV}}$) is equal to $2s^z$ ($l^z$) divided by the applied magnetic field. 
Accordingly, the magnitude of $\chi_{\rm{P}}$ ($\chi_{\rm{VV}}$) is affected by a spin (orbital) fluctuation.

Generally, the intra-orbital repulsion $U$ and the Hund coupling $J$ stabilize the
high-spin states and enhance spin susceptibility.
Accordingly, $\chi_{\rm{P}}$ grows with increasing $U$ or $J$.
On the other hand,  the inter-orbital repulsion $U'$ stabilizes the orbital moment by
prohibiting two electrons occupying different orbitals at the same site. The Hund coupling $J$ also destabilizes the
orbital moment by facilitating the simultaneous occupancy of different orbitals. As a result, $\chi_{\rm{VV}}$ grows with increasing $U'$, while it decreases with increasing $J$.

These contrastive correlation effects naturally lead to the different enhancement rates of $\chi_{\rm{VV}}$ and $\chi_{\rm{P}}$ noted above.
Although $\chi_{\rm{P}}$ is enhanced by a large intra-orbital repulsion $U$, the increase in $\chi_{\rm{VV}}$ is 
mainly brought about by a smaller inter-orbital repulsion $U'$.
Accordingly, $\chi_{\rm{VV}}/\chi^{(0)}_{VV}$ becomes relatively small compared with  $\chi_{\rm{P}}/\chi^{(0)}_P$. We plot the ratio $\chi_{\rm{VV}}/\chi_{\rm{P}}$ in Fig. \ref{Lambda0}(e) for several
values of $U'/U$, under the relation $U'=U-2J$.
Evidently, the correlation enhancement of $\chi_{\rm{VV}}$ is smaller
than that of $\chi_{\rm{P}}$ for a wide parameter range.

\begin{figure}[h]
\begin{center}
\includegraphics[width=0.5\textwidth]{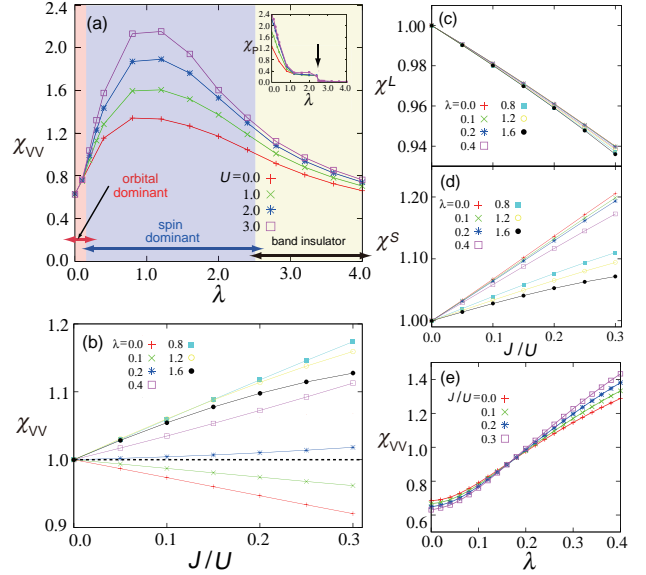}
\end{center}
\caption{\label{lambda} 
(Color online) 
(a) $\lambda$ dependences of $\chi_{\rm{VV}}$ for $U'/U=0.4$ and $J/U=0.3$. Inset: $\chi_{\rm{P}}$. The arrow shows the transition to the band insulator.
(b)-(d) $J/U$ dependences of $\chi_{\rm{VV}}$, $\chi^L$ and $\chi^S$ at $U=2.0$. They are normalized with
the values at $J=0$. (e)  $\chi_{\rm{VV}}$ at $U=2.0$ for small $\lambda$. $U'/U=0.4$ is fixed for (b)-(e). 
}
\end{figure}

Next, we will consider the correlation effects on $\chi_{\rm{P}}$ and $\chi_{\rm{VV}}$ under a finite spin-orbit coupling.
In Fig. \ref{lambda}(a), we plot the $\lambda$ dependence of $\chi_{\rm{VV}}$ for several $U$'s, with 
$U'/U=0.4$ and $J/U=0.3$ fixed. With this choice of $U'$ and $J$, $\chi_{\rm{VV}}$ is only slightly enhanced by inducing $U$ for $\lambda=0$, consistent with Fig. \ref{Lambda0}.
However, Fig. \ref{lambda}(a) clearly shows that $\chi_{\rm{VV}}$ increases with $U$ 
for moderate $\lambda$. $\chi_{\rm{VV}}/\chi_{\rm{P}}$ increases from $4.5$ to $6.3$ at $\lambda=1.2$ while sweeping $U$ from $0.0$ to $3.0$.

To elucidate the origin of this marked enhancement, we introduce the spin fluctuation $\chi^S$ and orbital fluctuation $\chi^L$ as
$\chi^S \equiv\frac{1}{N}\int\limits_0^{\beta}d\tau\langle S^z(\tau)S^z\rangle$ and
$\chi^L \equiv\frac{1}{N}\int\limits_0^{\beta}d\tau\langle L^z(\tau)L^z\rangle$, respectively,
with $S^{\alpha}(L^{\alpha})=\sum_{i,\eta\eta'}\langle\eta|s_i^{\alpha}(l_i^{\alpha})|\eta'\rangle c^{\dag}_{i\eta}c_{i\eta'}$.
In particular, $\chi^L$ corresponds to the fluctuation between $d_{yz}$ and $d_{zx}$ orbitals, which is essential to $\chi_{\rm{VV}}$ at $\lambda=0$.
We plot $\chi_{\rm{VV}}$ together with $\chi^S$ and $\chi^L$ in Figs. \ref{lambda}(b)-\ref{lambda}(d) with varying $J/U$.
As Figs. \ref{lambda}(c) and(d) show, the correlation effects on spin and orbital fluctuations are not sensitive to $\lambda$.
$\chi^S$ ($\chi^L$) monotonically increases (decreases) with $J$, consistent with the view that spin (orbital) fluctuation is enhanced (suppressed)
by Hund coupling.
In contrast, $\chi_{\rm{VV}}$ shows a non-monotonic $\lambda$ dependence.
For a small $\lambda$ ($\lambda\lesssim 0.2$), $\chi_{\rm{VV}}$ decreases with $J/U$, consistent with the case of $\lambda=0$, whereas, for an intermediate $\lambda$ ($\lambda\gtrsim 0.2$), $\chi_{\rm{VV}}$ grows with $J/U$\cite{largelambda}, as
is clearly shown in Fig. \ref{lambda}(e).

This non-monotonic behavior of $\chi_{\rm{VV}}$ can be ascribed to the mixing of spin and orbital degrees of freedom due to spin-orbit coupling.
Namely, the spin moment does not commute with the Hamiltonian for $\lambda\not=0$; hence, the spin degree of freedom also contributes to $\chi_{\rm{VV}}$. We define the spin- (orbital-) dominant region by the criterion $\chi_{\rm{VV}}[J/U=0.3]-\chi_{\rm{VV}}[J/U=0.0] > 0$ ($<0$),
and show the two regions in Fig. \ref{lambda}(a). 
Figure \ref{lambda}(a) clearly shows that $\chi_{\rm{VV}}$ is strongly enhanced in the spin-dominant region.
Namely, in the orbital-dominant region, $\chi_{\rm{VV}}$ is enhanced mainly by an inter-orbital electron interaction, whereas, in the spin-dominant region, a large intra-orbital electron interaction contributes to $\chi_{\rm{VV}}$, and
$\chi_{\rm{VV}}$ becomes strongly enhanced.
Our current analysis is based on the band structure of Sr$_2$RuO$_4$; however, we confirm that our results are generic in 
$t_{2g}$ systems by obtaining qualitatively the same behavior for other band structures.
The details of our analysis will be reported elsewhere.

Here, let us discuss our results, in association with experiments.
By adopting the band structure for Sr$_2$RuO$_4$, we revealed that $\chi_{\rm{VV}}/\chi_{\rm{P}}$ is $\sim 0.5$ in a non-interacting case. 
The electron interaction tends to make this ratio smaller for $\lambda\lesssim 0.2$.
Actually, the spin-orbit coupling of Sr$_2$RuO$_4$ is rather weak.
Therefore, it is reasonable to conclude that $\chi_{\rm{P}}$ dominates $\chi_{\rm{VV}}$ for Sr$_2$RuO$_4$, i.e., $\chi_{\rm{P}}$ has a dominant contribution to the Knight-shift signal.

On the other hand, for most Ir compounds, spin-orbit coupling is estimated to be fairly large.
In light of our analysis, $\chi_{\rm{VV}}$ is highly enhanced by electron correlation, if the spin-orbit coupling
is so large that the spin degree of freedom contributes to $\chi_{\rm{VV}}$.
This gives a possible clue to the anomalously large residual susceptibility of Ir compounds.
We note that some of the Ir compounds are considered as Mott insulators, which are outside the scope of our current analysis
based on a perturbative method.
Nevertheless, we consider that our mechanism is also relevant to the large residual susceptibility of such compounds.
It is an interesting future study to extend our analysis to the vicinity of metal-insulator transition.

In summary, we have studied the effects of electron correlation on the Pauli susceptibility $\chi_{\rm{P}}$ and the Van Vleck susceptibility $\chi_{\rm{VV}}$ on the basis of the multi-orbital Hubbard model.
We adopt DMFT combined with IPT, and calculate $\chi_{\rm{P}}$ and $\chi_{\rm{VV}}$, following the definitions introduced by Kontani and Yamada.
As a result, we found that the correlation enhancement of $\chi_{\rm{VV}}$ is rather small for a small $\lambda$. Meanwhile, a substantial increase is found for an intermediate $\lambda$, where the spin degree of freedom contributes to $\chi_{\rm{VV}}$.
 
We would like to thank H.\ Takagi, A.\ Yamamoto, and Y.\ Motome for fruitful discussions. This work is supported by KAKENHI (Nos. 21740242 and 21340090).


\begin{thebibliography}{99}
\bibitem{omitFock} 
To avoid the unphysical deformation of the Fermi surface, we omit Fock terms by assuming that these terms are included in $H_0$ as 
a renormalization of hopping integrals.
\bibitem{largelambda} We note that $\chi^S$ and $\chi^L$ show irregular $\lambda$ dependence for $\lambda\gtrsim 2.5$ (not shown), owing to the transition to a band insulator.
\end{thebibliography}
\end{document}